\documentclass{article}

\usepackage{PRIMEarxiv}

\usepackage{eso-pic}
\usepackage[utf8]{inputenc} 
\usepackage[T1]{fontenc}    
\usepackage{hyperref}       
\usepackage{url}            
\usepackage{booktabs}       
\usepackage{amsfonts}       
\usepackage{nicefrac}       
\usepackage{microtype}      
\usepackage{lipsum}
\usepackage{fancyhdr}       
\usepackage{graphicx}       
\graphicspath{{media/}}     
\usepackage{booktabs}
\usepackage{multirow}
\usepackage{graphicx}
\usepackage[table]{xcolor}
\usepackage{array}
\usepackage{caption}
\usepackage{natbib}
\usepackage[most]{tcolorbox}
\usepackage{listings}
\usepackage{wrapfig}

\newtcolorbox{CoTpromptbox}[1]{
    enhanced,
    breakable,
    colback=white,
    colframe=black!30,
    colbacktitle=black!78,
    coltitle=white,
    boxrule=0.5pt,
    arc=2pt,
    title={#1},
    fonttitle=\bfseries\small,
    fontupper=\normalfont\small,
    left=10pt,
    right=10pt,
    top=6pt,
    bottom=6pt,
    boxsep=2pt,
    before upper={\raggedright\setlength{\parindent}{0pt}\setlength{\parskip}{1pt}\setlength{\leftmargini}{1.5em}\setlength{\labelsep}{0.4em}},
    before skip=6pt,
    after skip=7pt
}
\newcommand{\promptheading}[1]{%
    \par\smallskip\noindent\textcolor{citeblue}{\bfseries #1}\par
}

\lstdefinestyle{CoTpromptstyle}{
    basicstyle=\ttfamily\footnotesize,
    columns=fullflexible,
    breaklines=true,
    breakatwhitespace=false,
    frame=single,
    framerule=0.3pt,
    rulecolor=\color{black!20},
    backgroundcolor=\color{black!3},
    xleftmargin=4pt,
    xrightmargin=4pt,
    framesep=3pt,
    aboveskip=2pt,
    belowskip=2pt
}
\newtcblisting{CoTpromptlisting}[1]{
    enhanced,
    breakable,
    listing only,
    listing options={style=CoTpromptstyle},
    colback=gray!3,
    colframe=black!55,
    colbacktitle=black!65,
    coltitle=white,
    boxrule=0.6pt,
    arc=2pt,
    title=\textbf{#1},
    fonttitle=\bfseries\small,
    left=6pt,
    right=6pt,
    top=5pt,
    bottom=5pt,
    boxsep=2pt,
    before skip=6pt,
    after skip=8pt
}

\definecolor{citeblue}{HTML}{335E96}
\hypersetup{
    colorlinks=true,
    citecolor=citeblue
}

\title{AntEmbed: Dual-Mode Multimodal Embedding Model for Efficient and Fine-Grained Retrieval
}

\author{
  Kehang Jia\textsuperscript{1,*},
  Teng Liu\textsuperscript{1,*},
  Juntao Li\textsuperscript{1,\textdagger},
  Hua Chen\textsuperscript{2},
  Jinxiong Chang\textsuperscript{2,\textdagger},
  Xiaobo Liang\textsuperscript{1},
  Min Zhang\textsuperscript{1,\textdagger}
  \\[0.08in]
  \textsuperscript{1}Soochow University
  \quad
  \textsuperscript{2}Ant Group
  \\[0.04in]
  \texttt{\{khjia,tliu2024\}@stu.suda.edu.cn}
  \quad
  \texttt{\{ljt,xbliang,minzhang\}@suda.edu.cn}
  \\
  \texttt{\{jinxiong.cjx,xuanji.ch\}@antgroup.com}
}

\begin{document}

\AddToShipoutPictureFG*{%
  \AtTextUpperLeft{%
    \put(0,\LenToUnit{-4mm}){%
      \includegraphics[width=0.18\textwidth]{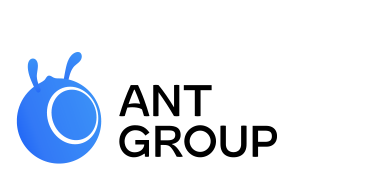}%
    }%
  }%
}

\maketitle

\begingroup
\renewcommand{\thefootnote}{\fnsymbol{footnote}}
\footnotetext[1]{Equal contribution. Work done during an internship at Ant Group.}
\footnotetext[2]{Corresponding authors.}
\endgroup

\AddToShipoutPictureFG*{%
  \AtTextUpperLeft{%
    \put(0,\LenToUnit{-4mm}){%
      \includegraphics[width=0.18\textwidth]{figures/ant-open-source.png}%
    }%
  }%
}

\maketitle

\definecolor{abstractblue}{HTML}{5F92D2}

\renewenvironment{abstract}
{
\begin{tcolorbox}[
    enhanced,
    colback=white,
    colframe=abstractblue,
    boxrule=0.8pt,
    arc=6pt,
    outer arc=6pt,
    left=10pt,
    right=10pt,
    top=8pt,
    bottom=8pt,
    before skip=6pt,
    after skip=10pt
]
\centerline{
    {\large\bfseries\scshape\textcolor{abstractblue}{Abstract}}
}
\vspace{4pt}
}
{
\end{tcolorbox}
}

\begin{abstract}

Discriminative embeddings directly encode original multimodal inputs into vector representations, offering low encoding overhead and efficient retrieval, and therefore serve as the primary representation paradigm for large-scale online retrieval. 
Generative embeddings, in contrast, use reasoning to organize query intent and relevance evidence, providing richer semantic representations for complex queries. 
To meet the demands of both large-scale low-latency retrieval and complex fine-grained matching, multimodal embedding model should not only support both representation capabilities, but also maintain their consistency within a unified representation space. 
To this end, we propose AntEmbed, a dual-mode multimodal embedding model that supports both efficient and fine-grained retrieval.
First, we construct retrieval-oriented reasoning content that explicitly captures query intent and confusion-aware fine-grained semantic information. 
We treat the original and reasoning-enhanced forms of each candidate as semantically equivalent views of the same instance and include both in the positive set for the corresponding query.
Through multi-positive contrastive learning, we establish cross-view matching constraints using a single embedding token, thereby aligning discriminative and generative embeddings within a unified embedding space.
Second, we introduce cross-mode self-distillation, which transfers fine-grained semantic information from reasoning-enhanced representations to original representations through candidate similarity-distribution distillation and query representation alignment. This enables discriminative embeddings to achieve stronger semantic modeling without requiring additional reasoning.
On the MMEB-v2 benchmark, AntEmbed-8B achieves a state-of-the-art overall score of 81.5 with its generative embeddings, with particularly strong performance on image retrieval and visual document retrieval.
\end{abstract}

\begin{figure*}[h]
  \centering
  \includegraphics[width=0.84\textwidth]{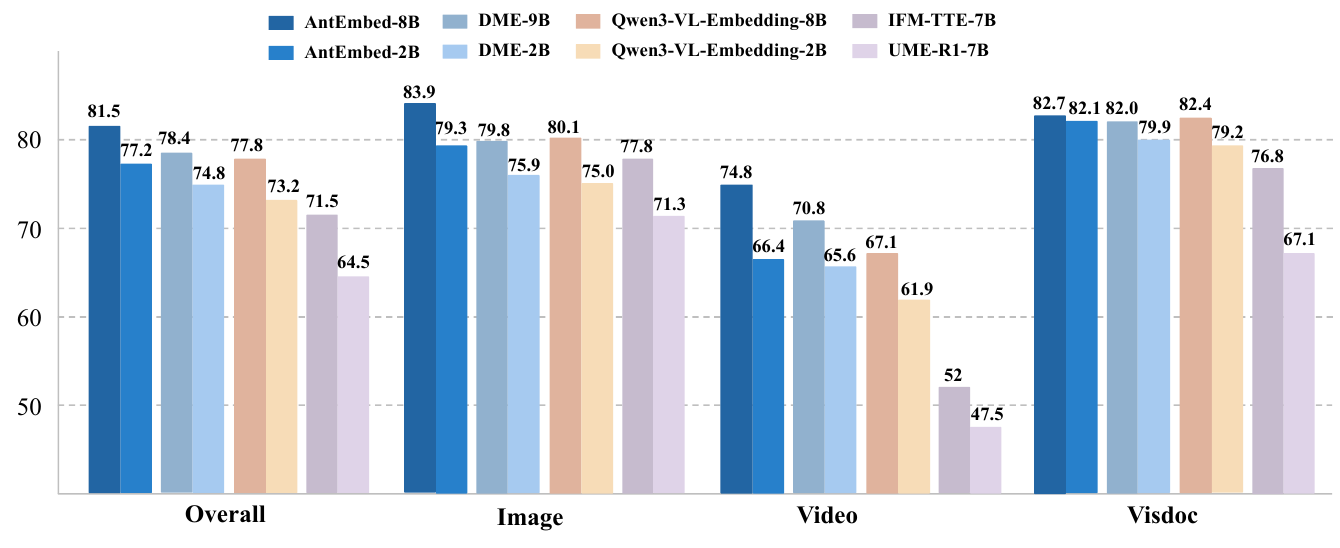}
  \caption{\textbf{Performance Comparison on MMEB-v2.} Overall and per-domain results on MMEB-v2. AntEmbed consistently outperforms competing generative and discriminative embedding models at comparable parameter scales.
}
  \label{fig:score}
\end{figure*}


\section{Introduction}

Multimodal large language models (MLLMs)\cite{achiam2023gpt,wu2023multimodal,liu2023visual}, with their stronger vision–language understanding and unified modeling of multimodal compositional inputs, have increasingly become the backbone of general-purpose multimodal retrievers.
Existing approaches\cite{zhang2024gme,li2026qwen3embedding} typically employ large-scale contrastive learning to directly encode original multimodal inputs into vectors, using the final hidden state of a predefined embedding token as the embedding representation. Since they do not generate intermediate content, these are referred to as discriminative embeddings\cite{lan2026ume}. 
Discriminative embeddings enable efficient encoding, but rely on direct input embedding, making it difficult to capture fine-grained attributes and the key semantic evidence underlying relevance judgments in complex inputs.

\begin{figure*}[t]
  \centering
  \includegraphics[width=0.94\textwidth]{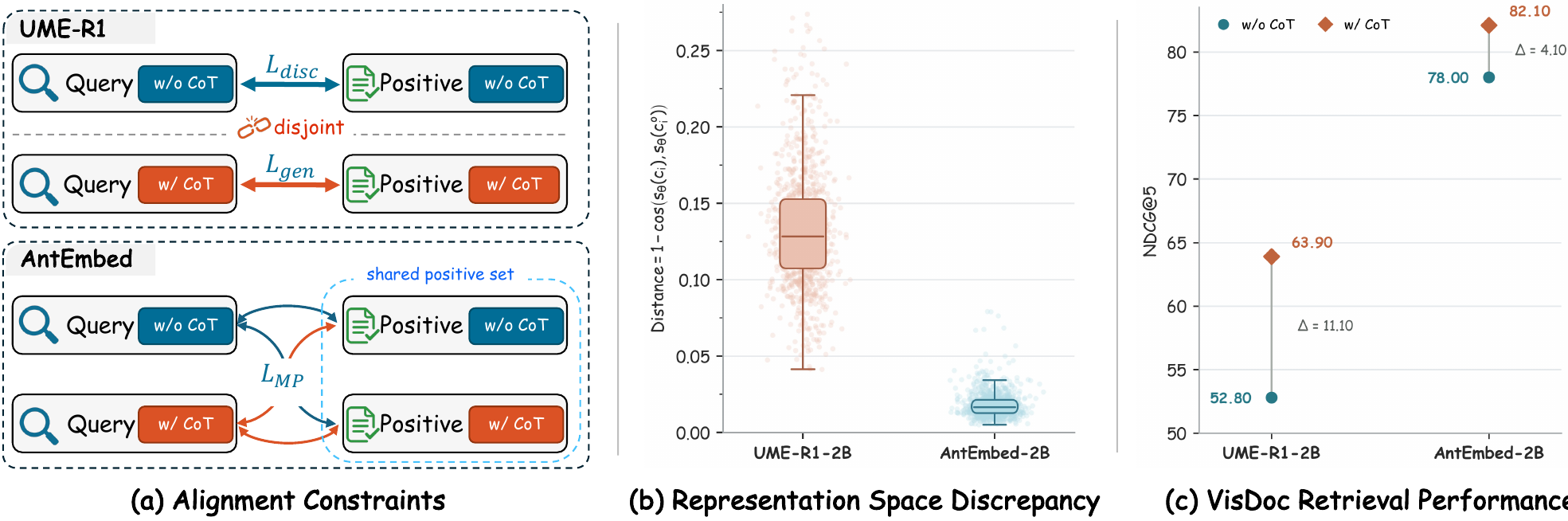}
  \caption{\textbf{Comparison between UME-R1 and AntEmbed.}
  UME-R1 optimizes the original and reasoning-enhanced inputs with separate
  within-view objectives, whereas AntEmbed treats them as semantically equivalent
  positive views and applies multi-positive contrastive learning to align their
representations in a unified embedding space.
  This produces a smaller representation gap between the two views and
  improves retrieval performance}
  \label{fig:motivation}
\end{figure*}

Recent studies\cite{cui2026think,liu2025reasoning,lan2026ume} have introduced Chain-of-Thought (CoT)\cite{wei2022chain} reasoning into the multimodal embedding, where intermediate reasoning processes explicitly organize query intent and relevance evidence to guide representation. The resulting representations are referred to as generative embedding.
Existing studies\cite{cui2026think,lan2026ume} show that generative embeddings can provide additional gains on complex and fine-grained multimodal tasks. However, they require an additional autoregressive reasoning process, which incurs substantial inference latency and computational overhead. 
This raises an important requirement: \textit{how to equip a single retriever with both the efficient discriminative embeddings and the fine-grained generative embeddings, enabling fast encoding and large-scale retrieval in standard scenarios while exploiting reasoning information to improve retrieval quality in complex scenarios.}
Methods such as UME-R1\cite{lan2026ume} address this requirement by introducing two independent embedding tokens, explicitly modeling discriminative and generative embeddings as separate embedding modes.

As shown in Figure~\ref{fig:motivation}(b), the representations of the same instance under the two modes exhibit low similarity, indicating a clear separation between their representation spaces. 
Under this design, \textbf{the representation is determined not only by its semantic content but also by the selected embedding mode.} 
Consequently, retrieval systems must predefine the embedding mode and maintain separate corpus representations for different modes, introducing substantial index storage overhead.
Notably, Figure~\ref{fig:motivation}(c) also shows a substantial performance gap between the generative and discriminative embeddings, suggesting that the reasoning information can effectively enhance representation quality. 
Meanwhile, discriminative embeddings remain the primary representation for low-latency online retrieval; improving their semantic modeling capability could reduce the system’s reliance on generative embeddings. 
This motivates us to investigate: \textbf{Can the reasoning information captured by generative embeddings be used to enhance discriminative embeddings, enabling stronger semantic modeling while retaining efficient retrieval?}

Based on the above observations, we propose AntEmbed, a dual-mode multimodal embedding model trained with a two-stage strategy.
AntEmbed aims to accommodate both discriminative and generative embeddings within a unified embedding space and establish consistent retrieval representations across views. 

Specifically, for each retrieval query, we treat the original and reasoning-enhanced forms of its positive as semantically equivalent views of the same instance and jointly include them in the query’s positive set.
The reasoning-enhanced view consists of the original input together with retrieval-oriented intent and confusion-aware fine-grained semantic information. 
In the first stage, the model uses a single predefined embedding token and multi-positive contrastive learning to jointly constrain the representations of original and reasoning-enhanced inputs, aligning discriminative and generative embeddings within a unified embedding space and establishing consistent retrieval representations across views.
In the second stage, we further use the reasoning-enhanced branch as the teacher and perform cross-mode self-distillation at both the candidate similarity distribution and query representation levels. This transfers the fine-grained semantic information obtained through reasoning to the discriminative embeddings, enabling stronger semantic modeling without additional inference.

Our main contributions are summarized as follows:

\begin{itemize}
\item We propose AntEmbed, a dual-mode multimodal embedding model that supports both efficient and fine-grained retrieval. AntEmbed treats original and reasoning-enhanced inputs as semantically equivalent views of the same instance and aligns them in a unified embedding space, enabling consistent retrieval across views.

\item We first treat the original and reasoning-enhanced candidate forms simultaneously as positives for each query and align their representations through multi-positive contrastive learning. We then introduce cross-mode self-distillation to transfer fine-grained semantic information from reasoning-enhanced representations to original representations, enhancing discriminative embeddings while reducing reliance on additional reasoning.

\item On the MMEB-v2 benchmark, AntEmbed-8B achieves state-of-the-art performance with its generative embeddings, while its discriminative embeddings remain competitive, outperforming some generative baselines.

\end{itemize}

\section{Related Work}
\label{sec:related_work}

\subsection{Multimodal Embedding Models}

Multimodal embedding models aim to map heterogeneous modalities into a shared representation space for unified semantic matching and multimodal retrieval. 
Early approaches, exemplified by vision–language contrastive models such as CLIP\cite{radford2021learning} and ALIGN\cite{jia2021scaling}, established cross-modal semantic alignment through large-scale image–text contrastive learning, laying the foundation for subsequent general-purpose multimodal embedding models.
However, these models typically rely on relatively independent modality-specific encoders, limiting their ability to handle complex inputs with arbitrary combinations of modalities. 
With the emergence of multimodal large language models (MLLMs)\cite{bai2025qwen3,li2024llava}, their stronger cross-modal understanding and joint modeling capabilities have enabled a new paradigm for constructing general-purpose multimodal embedding models. GME\cite{zhang2024gme}, VLM2Vec\cite{jiang2024vlm2vec}, and Qwen3-VL-Embedding\cite{li2026qwen3embedding} use MLLMs as backbones and adapt them through large-scale contrastive learning to support general-purpose multimodal representation and retrieval. 
Subsequent studies have further focused on complex inputs and fine-grained semantic matching. 
UniME\cite{gu2025breaking} and MM-Embed\cite{lin2025mm} introduce high-quality hard negatives to improve the model’s ability to distinguish fine-grained semantic differences among candidates. However, such instance-level supervision does not explicitly model why two instances are relevant or which fine-grained evidence should be preserved in their representations.

\subsection{Reasoning-Enhanced Generative Embedding}

Chain-of-Thought (CoT)\cite{wei2022chain} reasoning can make implicit intent, fine-grained attributes, and relevance evidence explicit, providing multimodal embedding models with richer semantic context. This has motivated growing interest in reasoning-enhanced generative embeddings. 
Existing generative embedding approaches broadly follow two architectural paradigms. The first is the decoupled reasoning–embedding architecture, exemplified by TTE\cite{cui2026think} and Embed-RL\cite{jiang2026embed}, separate the reasoner from the embedding model, allowing both components to scale independently and enabling stronger external reasoners to enrich semantic understanding. 
Alternatively, another paradigm\cite{liu2025reasoning} integrates reasoning and embedding into a unified multimodal model. While reducing communication overhead between separate modules, this design couples the two capabilities within a shared backbone, limiting their independent scalability. 
Regardless of the paradigm, generating explicit reasoning introduces additional computation compared with discriminative embedding. UME-R1\cite{lan2026ume} supports both discriminative and generative embeddings through two dedicated embedding tokens, allowing retrieval systems to select a representation mode according to task requirements and trade off efficiency against representational capacity. 
However, modeling the two modes separately can yield divergent representations and substantial retrieval performance gaps across modes. Maintaining separate candidate indexes for these modes further increases storage costs.

\subsection{Multi-Positive Contrastive Learning}

Standard contrastive learning typically assigns a single positive sample to each anchor and treats the remaining samples in the same batch as negatives. 
Multi-Positive contrastive learning instead associates each anchor with a set of semantically related positives, leveraging richer supervision by pulling multiple positives closer while pushing negatives apart. 
Some studies\cite{khosla2020supervised,wang2026training} incorporate multiple semantically related or similarly supervised samples into the positive set for contrastive learning.
StableRep\cite{tian2023stablerep} constructs positive sets from multiple images generated using the same text prompt, treating these images as mutual positives.
For generative embeddings, the original and reasoning-enhanced forms of a candidate correspond to the same retrieval instance and can therefore be treated as semantically equivalent views under different information conditions. 
However, this instance-level correspondence remains underexplored as a means of connecting discriminative and generative embeddings. We include both candidate views in the positive set for each corresponding query and use Multi-Positive contrastive learning to align representations across the two modes.

\section{Method}

\begin{figure*}[t]
  \centering
  \includegraphics[width=\textwidth]{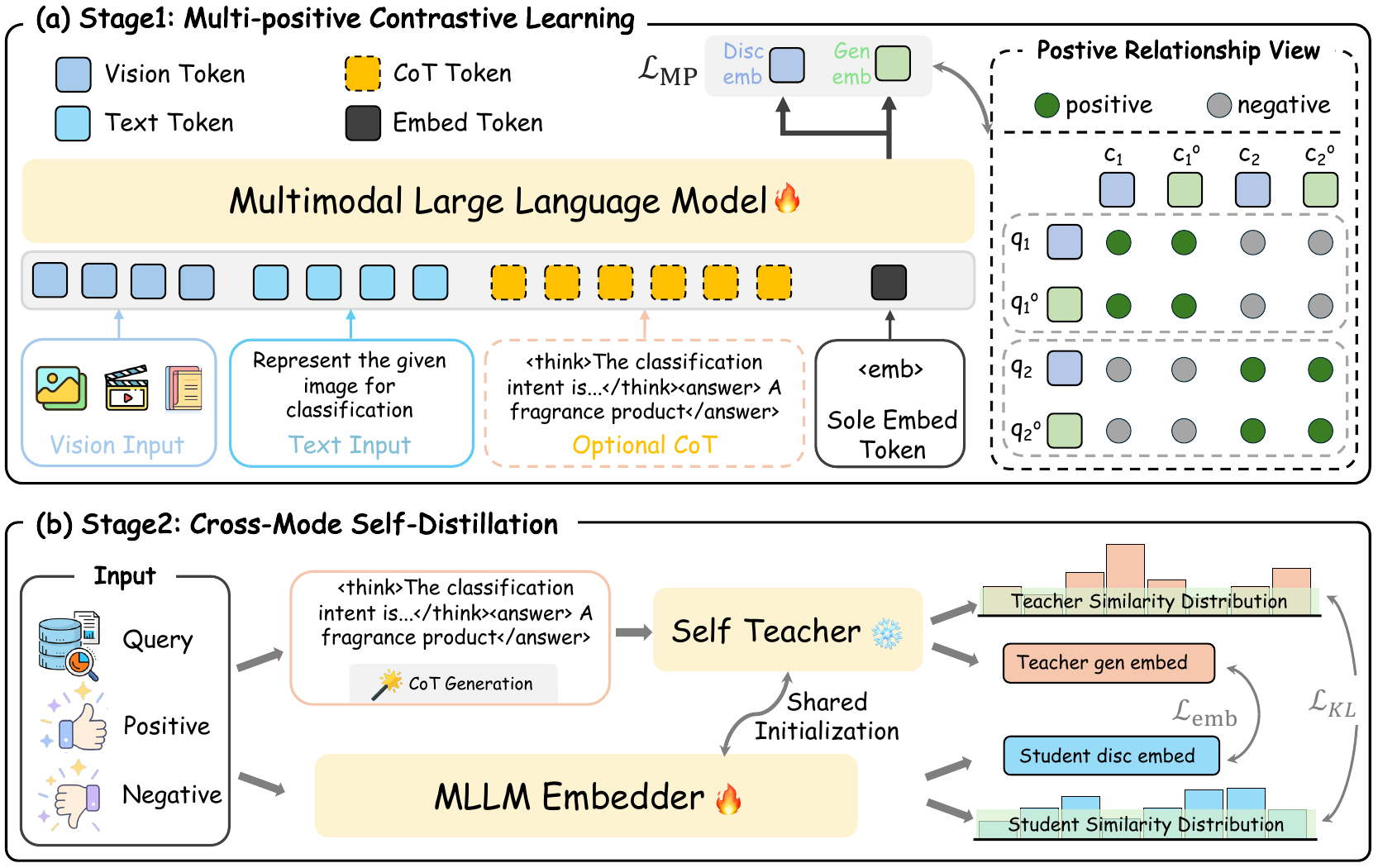}
  \caption{\textbf{Overview of AntEmbed.}
  In Stage 1, we construct original and reasoning-enhanced multimodal views and
  optimize them with multi-positive contrastive learning using a sole embedding
  token. In Stage 2, cross-mode self-distillation transfers the fine-grained semantic information obtained through reasoning to the discriminative embedding through candidate similarity-distribution distillation and query representation alignment.}
  \label{fig:framework}
\end{figure*}

This section first outlines the basic paradigms and training objectives of discriminative and generative embeddings (Sec.~\ref{sec:preliminary}). We then describe the CoT data construction process for AntEmbed (Sec.~\ref{sec:CoT_Construction}) and its two-stage training strategy. In the first stage, we employ multi-positive contrastive learning (Sec.~\ref{sec:multi-positive}) to jointly optimize the original and reasoning-enhanced forms as positive views of the same instance. In the second stage, we introduce cross-mode self-distillation (Sec.~\ref{sec:distillation}), using the reasoning-enhanced branch as the teacher and optimizing the original query branch through candidate similarity distribution distillation and query representation alignment.

\subsection{Preliminary}
\label{sec:preliminary}

Existing multimodal embedding models built upon MLLMs typically obtain input representations from the final hidden state of a predefined embedding token. Given a query $q$, a relevant positive candidate $c^{+}$, and a set of negative candidates $\mathcal{C} ^-=\left\{ c_{1}^{-},c_{2}^{-},...,c_{n}^{-} \right\}$, discriminative embedding methods directly encode the multimodal inputs and optimize the following InfoNCE objective\cite{oord2018representation} to pull $q$ closer to $c^{+}$ while pushing it away from the negative candidates in $\mathcal{C}^{-}$:
\begin{equation}
\label{eq:disc}
\mathcal{L} _{disc}=-\log \frac{\exp \left( \left( s_{\theta}\left( q \right) \cdot s_{\theta}\left( c^+ \right) \right) /\tau \right)}{\exp \left( \left( s_{\theta}\left( q \right) \cdot s_{\theta}\left( c^+ \right) \right) /\tau \right) +\sum\nolimits_{i=1}^N{\exp \left( \left( s_{\theta}\left( q \right) \cdot s_{\theta}\left( c_{i}^{-} \right) \right) /\tau \right)}}
\end{equation}
where $s_{\theta}$ denotes the embedding model with final normalization, and $\tau$ is the temperature hyperparameter. In contrast, generative embedding models first generate an intermediate reasoning trajectory and summary before producing the final representation. Their optimization objective can therefore be formulated as:
\begin{equation}
\mathcal{L} _{gen}=-\log \frac{\exp \left( \left( s_{\theta}\left( q^o \right) \cdot s_{\theta}\left( c^{+,o} \right) \right) /\tau \right)}{\exp \left( s_{\theta}\left( q^o \right) \cdot s_{\theta}\left( c^{+,o} \right) /\tau \right) +\sum\nolimits_{i=1}^N{\exp \left( \left( s_{\theta}\left( q^o \right) \cdot s_{\theta}\left( c_{i}^{-,o} \right) \right) /\tau \right)}}
\end{equation}
where $q^{o}$ and $c^{+,o}$ denote the reasoning-enhanced forms of the query and its corresponding positive candidate, respectively, obtained by concatenating the original input with the corresponding reasoning trajectory and summary.

\subsection{Confusion-Aware CoT Construction}
\label{sec:CoT_Construction}

 Unlike mathematical reasoning or code generation, multimodal embedding does not typically aim to derive a unique correct answer, but rather to determine the relevance between a query and a candidate. Therefore, the reasoning for retrieval should focus on relevance assessment, identifying the latent intent of the query, the semantic evidence supporting the relevance judgment, and the fine-grained attributes that distinguish semantically similar candidates.

Motivated by this observation, we introduce Confusion-Aware CoT Construction. Similar to TTE\cite{cui2026think}, we employ Qwen3-VL-30B-A3B-Instruct\footnote{\url{https://huggingface.co/Qwen/Qwen3-VL-30B-A3B-Instruct}}\cite{bai2025qwen3} as an external reasoner to generate reasoning trajectories for the training data. For a query $q$, we prompt the reasoner to focus on its retrieval intent and identify the key semantic evidence required to assess candidate relevance, rather than generating a specific answer. 
For a candidate $c$, semantic analysis based solely on its own content is insufficient to capture distinguishing information, as it lacks explicit comparisons with semantically similar candidates.
To address this issue, we first use the open-source multimodal embedding model Qwen3-VL-Embedding-8B\footnote{\url{https://huggingface.co/Qwen/Qwen3-VL-Embedding-8B}}\cite{li2026qwen3embedding} to retrieve the most similar candidates from the corpus, selecting the top-$K$ neighbors ($K$=5) to build the confusion set:

\begin{equation}
\mathcal{H} _K\left( c_i \right) =\mathrm{TopK}_{c_j\epsilon \mathcal{C}/\left\{ c_i \right\}}\left[ s_{\theta}\left( c_i \right) \cdot s_{\theta}\left( c_j \right) \right] 
\end{equation}

We then provide the candidate $c_i$ together with its confusion set to the reasoner, prompting it to identify the key attributes, relations, and visual or textual evidence that distinguish $c_i$ from these semantically similar candidates, thereby generating confusion-aware fine-grained semantic information.

\subsection{Multi-Positive Contrastive Learning}
\label{sec:multi-positive}

To avoid explicitly modeling discriminative and generative embeddings as two separate embedding modes, we extract both representations using a single predefined embedding token. If the original optimization objective is retained, the embedding space would establish the following matching relationships:

\begin{equation}
 q\leftrightarrow c^+;q^o\leftrightarrow c^{+,o} 
\end{equation}

Although both representations are extracted from the same embedding token, the model would still learn two independent within-mode matching objectives in the shared vector space. Consequently, the training process does not inherently impose semantic consistency constraints across different reasoning conditions.

To address this issue, we further treat different reasoning conditions of the same candidate as semantically equivalent positive views. 
Specifically, for an original query $q$ and its reasoning-enhanced counterpart $q^{o}$, we jointly include the original candidate $c$ and its reasoning-enhanced form $c^{+,o}$ as the positive set $\mathcal{P}(q) =\left\{ c^+,c^{+,o} \right\}$ for both queries.
A straightforward solution is to directly replace the single positive in Eq.~\ref{eq:disc} with the positive set and optimize the model using a multi-positive contrastive objective:

\begin{equation}
 \mathcal{L} =-\log \frac{\sum_{j\in \mathcal{P}}{\exp \left( \left( s_{\theta}\left( q \right) \cdot s_{\theta}\left( c_{j}^{+} \right) \right) /\tau \right)}}{\sum_{j\in \mathcal{P}}{\exp \left( \left( s_{\theta}\left( q \right) \cdot s_{\theta}\left( c_{j}^{+} \right) \right) /\tau \right)}+\sum_{i=1}^N{\exp \left( \left( s_{\theta}\left( q \right) \cdot s_{\theta}\left( c_{i}^{-} \right) \right) /\tau \right)}}
\end{equation}

However, this formulation aggregates the similarities of multiple positives into a single objective, emphasizing the overall probability assigned to the positive set rather than explicitly constraining each positive. This does not fully align with our objective: $c^+$ and $c^{+,o}$ represent the same candidate under different reasoning conditions, and both query conditions, $q$ and $q^{o}$, should maintain strong compatibility with each candidate view rather than relying exclusively on any single view. We therefore seek an objective that explicitly imposes a matching constraint on each positive sample. The contrastive objective is first reformulated from a pairwise ranking perspective:

\begin{equation}
 \mathcal{L} _{disc}=\log \left[ 1+\sum_{i=1}^N{\exp \left( \left( s_{\theta}\left( q \right) \cdot s_{\theta}\left( c_{i}^{-} \right) -s_{\theta}\left( q \right) \cdot s_{\theta}\left( c^+ \right) \right) /\tau \right)} \right] 
\end{equation}

This formulation makes the underlying ranking constraint explicit: each positive view is compared jointly against every negative candidate $c_i$.
Based on this formulation, for the two views of a query, $\mathcal{Q}=\left\{q,q^{o}\right\}$, and then extend the single positive to a positive set $\mathcal{P}$, requiring each positive to achieve a higher matching score than the negatives. 
This yields our final multi-positive training objective:

\begin{equation}
 \mathcal{L} _{MP}=\frac{1}{\left| \mathcal{Q}  \right|}\sum_{q\in \mathcal{Q}}{\log \left[ 1+\sum_{i=1}^N{\sum_{c_{j}^{+}\in \mathcal{P}(q)}{\exp \left( \left( s_{\theta}\left( q \right) \cdot s_{\theta}\left( c_{i}^{-} \right) -s_{\theta}\left( q \right) \cdot s_{\theta}\left( c_{j}^{+} \right) \right) /\tau \right)}} \right]}
\end{equation}

Rather than merely increasing the total probability mass assigned to the positive set, this objective explicitly preserves the relative matching relationship between each positive view and the negative candidates. 
Thus, both query conditions are matched against the same positive set:

\begin{equation}
q\leftrightarrow \left( c^+,c^{+,o} \right) ;q^o\leftrightarrow \left( c^+,c^{+,o} \right) 
\end{equation}

Accordingly, the original and reasoning-enhanced forms of each candidate are jointly incorporated into the optimization, and consistent matching constraints map representations across different views into a unified embedding space.

\subsection{Cross-Mode Self-Distillation}
\label{sec:distillation}

Although generative embeddings can capture richer semantic information through explicit reasoning, the additional autoregressive overhead makes them less suitable for routine encoding in online retrieval. 
To reduce reliance on computationally expensive generative embeddings, we further introduce cross-mode self-distillation, transferring semantic information from reasoning-enhanced representations to discriminative embeddings, thereby enhancing their semantic modeling capability.

Specifically, we initialize both the teacher and student branches from the first-stage model, feeding them with reasoning-enhanced and original inputs, respectively, and update only the student branch.
We then perform distillation at two levels: candidate similarity distributions and query representation alignment.
We first use the candidate similarity distribution produced by the reasoning-enhanced branch as a soft target to supervise the original query branch. The candidate similarity distribution of the student branch is defined as:
\begin{equation}
p^S\left( i \right|q) =\mathrm{softmax}_i\left( \frac{s_{\theta _S}\left( q \right) \cdot \mathrm{sg}\left( s_{\theta _S}\left( c_{i}^{\mathrm{o}} \right) \right)}{\tau _{\mathrm{KL}}} \right) 
\end{equation}
The definition of similarity distribution for teacher branches is as follows:
\begin{equation}
p^T\left( i \right|q^o) =\mathrm{softmax}_i\left( \frac{\mathrm{sg}\left( s_{\theta _T}\left( q^o \right) \right) \cdot \operatorname{sg}\left( s_{\theta _T}\left( c_{i}^{\mathrm{o}} \right) \right)}{\tau _{\mathrm{KL}}} \right) 
\end{equation}
where $s_{\theta_T}$ and $s_{\theta_S}$ denote the embedding encoders of the teacher and student branches, respectively. $\tau_{\mathrm{KL}}$ denotes the distillation temperature, and $\operatorname{sg}(\cdot)$ denotes the stop-gradient operation. We then use the teacher distribution to supervise the student distribution and define the similarity distribution distillation loss as:
\begin{equation}
\mathcal{L} _{\mathrm{KL}}=\sum_{i=1}^N{p^T\left( i|q^o \right) \log \frac{p^T\left( i|q^o \right)}{p^S\left( i|q \right)}}
\end{equation}

This loss not only conveys positive matching information but also preserves the teacher’s similarity distribution over the candidates, allowing discriminative embeddings to capture more fine-grained relevance information among candidates.

However, matching the candidate similarity distribution primarily constrains the relative similarity between queries and candidates and does not directly enforce consistency between discriminative and generative embeddings in the embedding space. To further transfer information from the reasoning-enhanced representations to the discriminative embeddings, we introduce a query representation alignment loss:

\begin{equation}
\mathcal{L} _{emb}=\frac{1}{2}\left\| s_{\theta _S}\left( q \right) -\operatorname{sg}\left( s_{\theta _T}\left( q^o \right) \right) \right\| _{2}^{2}
\end{equation}

This objective directly enforces consistency between discriminative and generative embeddings by aligning the query representation with its reasoning-enhanced counterpart, thereby improving the semantic modeling capability of the discriminative embedding.
The final cross-mode self-distillation objective combines the multi-positive contrastive loss, similarity distribution distillation loss, and query representation alignment loss:

\begin{equation}
\mathcal{L} =\mathcal{L} _{MP}+\lambda _{\mathrm{KL}}\mathcal{L} _{\mathrm{KL}}+\lambda _{emb}\mathcal{L} _{emb}
\end{equation}

Through cross-mode self-distillation, discriminative embeddings learn finer-grained candidate relevance signals and semantic representations, enabling stronger modeling of complex semantics without incurring additional overhead.

\section{Experiments}

\subsection{Experimental Setup}

\textbf{Training Details.} We use Qwen3-VL-2B-Instruct\cite{bai2025qwen3} and Qwen3-VL-8B-Instruct as the backbone models for training the AntEmbed series. In the first stage, we train on the MMEB-v2 training set together with other publicly available multimodal retrieval datasets, and using Low-Rank Adaptation (LoRA) with rank ($r$=16, $\alpha$=32). The contrastive learning temperature is set to 0.02. We employ GradCache\cite{gao2021scaling} to scale the global batch size to 2048. The learning rate is set to $1\times10^{-4}$, and the model is trained for one epoch, corresponding to approximately 5,000 steps.
In the second stage, we continue training from the first-stage checkpoint using a curated subset of image and visual document data. The KL distillation temperature is set to 0.05, while the loss weights for similarity distribution distillation and query representation alignment are set to $\lambda_{\mathrm{KL}}=1.0$ and $\lambda_{\mathrm{emb}}=0.5$.

\textbf{Benchmarks and Baselines.}We use MMEB-V2\cite{meng2025vlm2vec} as our primary benchmark and compare AntEmbed with existing multimodal embedding models. We further extend the evaluation to MMEB-V3\cite{huang2026mmebv3}. MMEB-V2 covers diverse retrieval tasks across image, video, and visual document modalities. We report Hit@1 for image and video retrieval tasks, and NDCG@5\cite{jarvelin2002cumulated} for visual document tasks.
We compare against two categories of baselines. Discriminative embedding baselines include VLM2Vec\cite{meng2025vlm2vec}, GME\cite{zhang2024gme}, and Qwen3-VL-Embedding\cite{li2026qwen3embedding}. Generative embedding baselines include TTE$_t$\cite{cui2026think}, which uses Qwen2.5-VL-72B\cite{Qwen2.5-VL} as the external reasoner, as well as Embed-RL\cite{jiang2026embed} and UME-R1\cite{lan2026ume}.

\subsection{Main Results}

\begin{table*}[t]
\centering
\setlength{\tabcolsep}{5.2pt}
\renewcommand{\arraystretch}{1.32}

\caption{Comparison of performance between baselines and AntEmbed on MMEB-V2. Baseline results are taken from the official MMEB Leaderboard. \textbf{CLS}: classification; \textbf{QA}: question answering; \textbf{RET}: retrieval; \textbf{GD}: visual grounding; \textbf{VDR}: ViDoRe; \textbf{VR}: VisRAG; \textbf{OOD}: out-of-domain.}
\label{tab:main_result}

\resizebox{\textwidth}{!}{%
\begin{tabular}{l*{16}{c}}
\toprule

\multirow{2}{*}{Model}
& \multicolumn{5}{c}{\textbf{Image}}
& \multicolumn{5}{c}{\textbf{Video}}
& \multicolumn{5}{c}{\textbf{VisDoc}}
& \multirow{2}{*}{All}
\\

\cmidrule(lr){2-6}
\cmidrule(lr){7-11}
\cmidrule(lr){12-16}

& CLS & QA & RET & GD & Overall
& CLS & QA & RET & MRET & Overall
& VDRv1 & VDRv2 & VR & OOD & Overall
\\

\midrule

\# datasets
& 10 & 10 & 12 & 4 & 36
& 5 & 5 & 5 & 3 & 18
& 10 & 4 & 6 & 4 & 24
& 78
\\

\midrule

\rowcolor{gray!20}
\multicolumn{17}{c}{\textit{Baseline Models}}
\\

\midrule

VLM2Vec-8B\cite{jiang2024vlm2vec}
 & 62.7 & 56.9 & 69.4 & 82.2 & 65.5 & 39.1 & 30.0 & 29.0 & 38.9 & 33.7 & 56.9 & 9.4 & 59.1 & 54.0 & 49.1 & 53.1
\\

VLM2Vec-V2-2B\cite{meng2025vlm2vec}  & 62.9 & 56.3 & 69.5 & 77.3 & 64.9 & 39.3 & 34.3 & 28.8 & 36.8 & 34.6 & 75.5 & 44.9 & 79.4 & 62.2 & 69.2 & 59.2
\\

GME-2B\cite{zhang2024gme}
& 54.4 & 29.9 & 66.9 & 55.5 & 51.9
& 34.9 & 42.0 & 25.6 & 32.4 & 33.9
& 86.1 & 54.0 & 82.5 & 43.1 & 72.7
& 54.1
\\

GME-7B\cite{zhang2024gme}
& 57.7 & 34.7 & 71.2 & 59.3 & 56.0
& 37.4 & 50.4 & 28.4 & 37.0 & 38.4
& \textbf{89.4} & 55.6 & 85.0 & 44.4 & 75.2
& 57.8
\\

Qwen3-VL-Embedding-2B\cite{li2026qwen3embedding} & 70.3 & 74.3 & 74.8 & 88.5 & 75.0 & 71.9 & 64.9 & 53.9 & 53.3 & 61.9 & 84.4 & 65.3 & 86.4 & 69.4 & 79.2 & 73.2
\\

Qwen3-VL-Embedding-8B\cite{li2026qwen3embedding} & 74.2 & 81.1 & 80.2 & 92.3 & 80.1 & 78.4 & 71.0 & 58.7 & 56.1 & 67.1 & 87.2 & \textbf{69.9} & 88.7 & 73.3 & 82.4 & 77.8
\\

Embed-RL-4B\cite{jiang2026embed} & 63.7 & 70.5 & 71.3 & 91.4 & 70.1 & 57.6 & 58.4 & 45.1 & 49.5 & 53.0 & 80.2 & 53.4 & 84.9 & 67.1 & 74.7 & 68.1
\\

UME-R1-2B\cite{lan2026ume}
& 64.8 & 62.8 & 67.6 & 77.2 & 66.6
& 44.3 & 51.0 & 32.9 & 39.7 & 42.2
& 72.4 & 46.2 & 79.2 & 37.2 & 63.9
& 60.1
\\

UME-R1-7B\cite{lan2026ume}
& 67.1 & 69.2 & 71.9 & 84.9 & 71.3
& 48.6 & 60.7 & 38.2 & 39.3 & 47.5
& 75.7 & 50.5 & 83.7 & 37.6 & 67.1
& 64.5
\\

TTE$_t$-2B\cite{cui2026think}
& 76.6 & 76.8 & 71.5 & 87.2 & 76.1
& 56.1 & 65.3 & 34.1 & 33.8 & 47.9
& 81.1 & 62.4 & 84.7 & 43.2 & 72.6
& 68.6
\\

TTE$_t$-7B\cite{cui2026think}
& 76.7 & 78.6 & 74.3 & 89.3 & 77.8
& 57.5 & 68.2 & 38.0 & 39.3 & 52.0
& 83.7 & 63.6 & 91.4 & 50.6 & 76.8
& 71.5
\\

DME-2B\cite{chen2026douyin} & 68.6 & 76.2 & 75.6 & 94.3 & 75.9 & 84.5 & 61.9 & 55.5 & 57.4 & 65.6 & 87.0 & 56.2 & 90.3 & 70.0 & 79.9 & 74.8
\\

DME-9B\cite{chen2026douyin} & 74.5 & 80.9 & 78.2 & \textbf{94.6} & 79.8 & \textbf{87.7} & 71.0 & 61.0 & 58.5 & 70.8 & 87.6 & 57.8 & \textbf{94.5} & \textbf{73.5} & 82.0 & 78.4
\\

\midrule

\rowcolor{gray!20}
\multicolumn{17}{c}{\textit{Ours}}
\\

\midrule

AntEmbed-2B (base)
& 70.2 & 70.3 & 73.0 & 89.4 & 73.3
& 61.3 & 60.8 & 49.1 & 36.1 & 53.6
& 84.0 & 58.2 & 87.5 & 68.4 & 78.0
& 70.2
\\

AntEmbed-2B (CoT)
& 78.9 & 77.2 & 77.8 & 90.1 & 79.3
& 72.9 & 66.7 & 66.7 & 54.8 & 66.4
& 88.7 & 64.4 & 90.2 & 71.6 & 82.1
& 77.2
\\

\midrule

AntEmbed-8B (base)
& 71.5 & 75.9 & 75.8 & 91.8 & 76.4
& 62.8 & 65.9 & 54.4 & 40.4 & 57.6
& 86.1 & 59.0 & 89.7 & 70.8 & 79.9
& 73.1
\\

AntEmbed-8B (CoT)
& \textbf{83.7} & \textbf{81.2} & \textbf{83.5} & 93.0 & \textbf{83.9}
& 79.8 & \textbf{74.2} & \textbf{76.5} & \textbf{64.4} & \textbf{74.8}
& 89.2 & 64.3 & 90.9 & 72.5 & \textbf{82.7}
& \textbf{81.5}
\\

\bottomrule
\end{tabular}%
}
\end{table*}

We construct a single generative embedding index for the MMEB-V2 candidate corpus and evaluate AntEmbed under two query modes: base (discriminative encoding only) and CoT (with CoT generation). 
As shown in Table 1, \textbf{(1)} AntEmbed achieves strong and consistent retrieval performance under both modes. The CoT mode consistently outperforms the base mode across image, video, and visual-document tasks, improving the overall score of AntEmbed-2B from 70.2 to 77.2. The improvements on image, visual-document, and video tasks are 6.0, 4.1, and 12.8 points, respectively. 
\textbf{(2)} With CoT generation, AntEmbed-8B achieves the best overall score of 81.5, surpassing the strongest baseline, DME-9B, by 3.1 points and Qwen3-VL-Embedding-8B by 3.7 points. It also obtains the highest category-level scores on Image (83.9), Video (74.8), and VisDoc (82.7), demonstrating strong and consistent retrieval performance across modalities. These results show that reasoning-enhanced embeddings are particularly effective for fine-grained multimodal retrieval.
\textbf{(3)} Even in the base mode, AntEmbed-2B still achieves an overall score of 70.2. Among discriminative embedding baselines, its performance is second only to Qwen3-VL-Embedding-2B (73.2).
Notably, although the base mode does not require additional query reasoning, it still outperforms generative embedding models such as TTE$_t$-2B (68.6) and Embed-RL-4B (68.1). These results demonstrate that AntEmbed can retain the efficiency of discriminative encoding while achieving competitive multimodal retrieval performance.

\subsection{Ablation Study}

\textbf{Generalization Evaluation on MMEB-v3.}
To further evaluate the generalization capability of AntEmbed, we conduct an extended evaluation on MMEB-v3\cite{huang2026mmebv3}. 
Building upon MMEB-v2, this benchmark further introduces agent-based tasks, audio tasks, and more challenging text retrieval tasks, covering scenarios such as tool retrieval, GUI control, agent memory retrieval, instruction following, reasoning, long-context retrieval, and multi-condition retrieval, with a total of 190 evaluation tasks. 
Since AntEmbed does not support audio inputs, we exclude audio tasks from the evaluation; consequently, the scores of the 11 audio tasks are set to 0 when computing the V3-All score. 
For the newly introduced tasks in MMEB-v3, AntEmbed is evaluated under the reasoning-enhanced condition, while the baseline results are taken from those reported in the original MMEB-v3 paper. 
As shown in Table~\ref{tab:abl_mmebv3}, AntEmbed-8B achieves the highest V3-All score of 55.6 among all compared models. It also ranks first in both the overall Text and Agent categories, with scores of 42.8 and 40.4, respectively. These results further demonstrate the strong generalization capability of AntEmbed across diverse retrieval tasks.


\begin{table*}[t]
\centering
\setlength{\tabcolsep}{4pt}
\renewcommand{\arraystretch}{1.15}

\caption{Evaluation results on MMEB-v3. We follow the evaluation metrics adopted in MMEB-v3, using NDCG@5 for text and visdoc tasks, and Hit@1 for all other tasks.
RR: reasoning retrieval; IF: instruction following; LC: long-context retrieval; MC: multi-condition retrieval; GR: general retrieval; Memory: agent memory retrieval. Tool: tool retrieval; GUI: GUI control.}
\label{tab:abl_mmebv3}

\resizebox{\textwidth}{!}{%
\begin{tabular}{l c c c c c c c c c c c c}
\toprule

\multirow{3}{*}{\textsf{\textbf{Model}}}
&
\multirow{3}{*}{\textsf{\textbf{V3-All}}}
&
\multicolumn{6}{c}{\textsf{\textbf{Text}}}
&
\multicolumn{4}{c}{\textsf{\textbf{Agent}}}
&
\multirow{3}{*}{\textsf{\textbf{MCMR}}}
\\

\cmidrule(lr){3-8}
\cmidrule(lr){9-12}

&
&
\textsf{\textbf{Overall}}
&
\textsf{\textbf{RR}}
&
\textsf{\textbf{IF}}
&
\textsf{\textbf{LC}}
&
\textsf{\textbf{MC}}
&
\textsf{\textbf{GR}}
&
\textsf{\textbf{Overall}}
&
\textsf{\textbf{Tool}}
&
\textsf{\textbf{GUI}}
&
\textsf{\textbf{Memory}}
&
\\

\midrule

\multicolumn{1}{l}{\# datasets}
&
190
&
53 & 20 & 9 & 6 & 5 & 13
&
47 & 35 & 8 & 4
&
1
\\

\midrule

VLM2Vec-7B\cite{jiang2024vlm2vec}
& 30.8
& 22.2 & 7.2 & 28.1 & 5.9 & 40.9 & 41.7
& 19.7 & 19.8 & 21.4 & 15.9
& 0.9
\\

VLM2Vec-V2-2B\cite{meng2025vlm2vec}
& 38.3
& 24.5 & 7.8 & 29.2 & 11.5 & 50.3 & 41.2
& 28.7 & 27.6 & 36.2 & 23.3
& 4.1
\\

GME-7B\cite{zhang2024gme}
& 43.0
& 37.1 & 12.5 & 52.4 & 17.8 & 59.0 & \textbf{62.5}
& 35.6 & 39.0 & 30.0 & 17.1
& 27.3
\\

Qwen3-VL-Embedding-2B\cite{li2026qwen3embedding}
& 50.9
& 39.2 & 16.6 & 40.6 & 53.1 & 61.2 & 58.0
& 39.3 & \textbf{42.6} & 30.4 & \textbf{28.4}
& 42.0
\\

Qwen3-VL-Embedding-8B\cite{li2026qwen3embedding}
& 49.9
& 42.5 & 18.2 & \textbf{44.8} & 58.0 & \textbf{61.2} & 61.2
& 38.4 & 41.3 & 33.5 & 22.8
& 38.0
\\

\midrule
\rowcolor{gray!15}
AntEmbed-2B
& 52.5
& 40.8 & 18.2 & 43.3 & 64.3 & 58.7 & 56.2
& 37.2 & 37.7 & 40.1 & 27.1
& 36.8
\\

\rowcolor{gray!15}
AntEmbed-8B
& \textbf{55.6}
& \textbf{42.8} & \textbf{21.5} & 44.6 & \textbf{67.4} & 57.3 & 57.4
& \textbf{40.4} & 41.2 & \textbf{42.6} & 28.3
& \textbf{42.1}
\\

\bottomrule
\end{tabular}%
}

\end{table*}

\textbf{Effectiveness of Multi-Positive Training.} 
We first train two single-condition baselines using only original data or only CoT data. 
As shown in Table~\ref{tab:abl_multipos}, Base-only and CoT-only achieve average scores of 68.7 and 73.6, respectively, under their corresponding evaluation modes. 
To support both discriminative embedding and generative embedding within a shared embedding space, we first consider two straightforward approaches.
The first jointly trains the model on original and CoT data using a single predefined embedding token. The second trains separate LoRA adapters for the two conditions and merges their parameters (LoRA Merge). 
Both approaches perform comparably to the corresponding single-condition baselines, indicating that neither data mixing nor parameter merging alone yields substantial cross-condition gains. 
In contrast, our multi-positive training achieves average scores of 73.3 in Base mode and 79.5 in CoT mode, outperforming the other training strategies in both settings.

\begin{table}[t]
\centering
\setlength{\tabcolsep}{7pt}
\renewcommand{\arraystretch}{1.4}

\caption{Comparison of different training strategies with training tailored to either original or CoT data alone. We report the results of both embedding modes on the Image modality of MMEB-V2.
}
\label{tab:abl_multipos}

\begin{tabular}{l*{5}{c}p{0.8em}*{5}{c}}
\toprule

\multirow{2}{*}{\textsf{\textbf{Model}}}
&
\multicolumn{5}{c}{\textsf{\textbf{Base-mode Eval}}}
&
&
\multicolumn{5}{c}{\textsf{\textbf{CoT-mode Eval}}}
\\

\cmidrule(lr){2-6}
\cmidrule(lr){8-12}

&
\textsf{\textbf{CLS}}
&
\textsf{\textbf{QA}}
&
\textsf{\textbf{Ret}}
&
\textsf{\textbf{GD}}
&
\textsf{\textbf{AVG.}}
&
&
\textsf{\textbf{CLS}}
&
\textsf{\textbf{QA}}
&
\textsf{\textbf{Ret}}
&
\textsf{\textbf{GD}}
&
\textsf{\textbf{AVG.}}
\\

\midrule

\multicolumn{12}{c}{\textit{Data Composition}}
\\



Base-only
& 63.9 & 64.3 & 70.4 & 86.4 & 68.7 & & — & — & — & — & —
\\

CoT-only
& — & — & — & — & — & & 72.1 & 72.4 & 71.1 & 87.8 & 73.6
\\

\midrule

\multicolumn{12}{c}{\textit{Training Strategy}}
\\

Joint Training
& 66.5 & 64.6 & 68.7 & 85.6 & 68.8 & & 69.8 & 57.7 & 67.1 & 83.1 & 67.0
\\

LoRA Merge
& 65.97 & 65.0 & 69.3 & 85.4 & 69.0 & & 73.0 & 72.2 & 71.2 & 87.8 & 73.8
\\

\midrule

\rowcolor{gray!15}

Multi-Positive Training
& \textbf{70.2} & \textbf{70.3} & \textbf{73.0} & \textbf{89.4} & \textbf{73.3} & & \textbf{79.3} & \textbf{79.4} & \textbf{76.1} & \textbf{90.3} & \textbf{79.5}
\\

\bottomrule
\end{tabular}
\end{table}

\begin{table*}[t]
\centering

\begin{minipage}[t]{0.49\textwidth}
\vspace{0pt}
\centering

\caption{Ablation of the two-stage training strategy and the contribution of the two distillation objectives.}
\label{tab:ablation_two_stage}

\renewcommand{\arraystretch}{1.3}
\setlength{\tabcolsep}{3pt}
\scriptsize

\resizebox{\linewidth}{!}{%
\begin{tabular}{ccccccc}
\toprule
\textbf{Configuration}
&
\textbf{$\lambda_{\mathrm{KL}}$}
&
\textbf{$\lambda_{\mathrm{emb}}$}
&
\textbf{Image}
&
\textbf{Video}
&
\textbf{VisDoc}
&
\textbf{All}
\\
\midrule
Stage 1 & -- & -- & 72.0 & 52.5 & 77.2 & 69.1 \\
Stage 2 & 1 & 0 & 72.5 & 52.7 & 77.4 & 69.4 \\
Stage 2 & 0 & 1 & 72.6 & 53.1 & 77.2 & 69.5 \\
Stage 2 & 1 & 1 & 73.0 & 53.1 & 77.2 & 69.7 \\
\rowcolor{gray!15}
Stage 2 & 1 & 0.5 & \textbf{73.3} & \textbf{53.6}
& \textbf{78.0} & \textbf{70.2} \\
\bottomrule
\end{tabular}%
}

\end{minipage}
\hfill
\begin{minipage}[t]{0.49\textwidth}
\vspace{0pt}
\centering

\caption{Effect of different reasoners on generative embedding performance. ``w/o Reasoner'' denotes evaluation with base mode.}
\label{tab:ablation_reasoner}

\renewcommand{\arraystretch}{1.3}
\setlength{\tabcolsep}{3pt}
\scriptsize

\resizebox{\linewidth}{!}{%
\begin{tabular}{lccccc}
\toprule
\textbf{Reasoner}
&
\textbf{CLS}
&
\textbf{QA}
&
\textbf{Ret}
&
\textbf{GD}
&
\textbf{AVG.}
\\
\midrule
w/o Reasoner
& 70.2 & 70.3 & 73.0 & 89.4 & 73.3 \\

Qwen3-VL-8B-Instruct (SFT)
& 77.8 & 75.5 & 76.8 & 89.0 & 78.1 \\

Qwen3-VL-30B-A3B-Instruct
& 78.9 & 77.2 & \textbf{77.8} & 90.1 & 79.3 \\

Gemini-3.1-Pro-Preview
& \textbf{79.2} & \textbf{80.1} & 77.7 & \textbf{90.2} & \textbf{80.2} \\

\bottomrule
\end{tabular}%
}

\end{minipage}

\end{table*}

\textbf{Two-stage Training.} To validate the effectiveness of the two-stage training strategy, we compare a model trained only with the first-stage multi-positive contrastive learning against one further optimized with cross-mode self-distillation. 
As shown in Table~\ref{tab:ablation_two_stage}, the first-stage model achieves an average score of 69.1 across image, visual document, and video tasks, while incorporating the second-stage cross-mode self-distillation improves the overall performance to 70.2, further strengthening the representation capability of the discriminative embedding. 
We further ablate the two distillation objectives by varying $\lambda_{\mathrm{KL}}$ and $\lambda_{\mathrm{emb}}$ in cross-mode self-distillation.
As shown in Table~\ref{tab:ablation_two_stage}, either similarity distribution distillation or query representation alignment consistently improves over the first-stage model, while jointly applying both objectives achieves the best performance of 70.2.
These results demonstrate that similarity distribution distillation and query representation alignment provide complementary supervision at the levels of retrieval relations and representation space, respectively, and that their combination more effectively transfers reasoning-enhanced information to discriminative embeddings.

\textbf{Effect of Reasoner.}
Our decoupled reasoning–embedding architecture allows flexible use of external reasoning models to enhance multimodal semantic understanding.
We therefore investigate how the capability of external reasoners affects the quality of generative embeddings in AntEmbed by constructing evaluation-time reasoning chains with different reasoners. Specifically, we consider three external reasoners: a Qwen3-VL-8B-Instruct model fine-tuned with the training data, Qwen3-VL-30B-A3B-Instruct used to construct our training data, and Gemini-3.1-Pro-Preview. 
As shown in Table~\ref{tab:ablation_reasoner}, Qwen3-VL-8B-Instruct (SFT) improves the average score from 73.3 in the base mode to 78.1, yielding a gain of 4.8 points. This result demonstrates that external reasoning can substantially enhance retrieval performance, even with a relatively less capable reasoner. Moreover, stronger reasoners consistently lead to better retrieval performance: Qwen3-VL-30B-A3B-Instruct and Gemini-3.1-Pro-Preview further improve the average score to 79.3 and 80.2, respectively. These results demonstrate that AntEmbed can flexibly leverage external reasoners with different capabilities to further enhance retrieval performance.

\section{Conclusion and Future Work}

In this work, we proposed AntEmbed, a dual-mode multimodal embedding model that supports both efficient discriminative retrieval and reasoning-enhanced fine-grained retrieval. AntEmbed uses a two-stage training strategy to bring original and reasoning-enhanced inputs into a unified embedding space. Multi-positive contrastive learning aligns the two views of each instance, while cross-mode self-distillation transfers fine-grained relevance information from reasoning-enhanced representations to discriminative representations. On MMEB-v2, AntEmbed-8B achieves an overall score of 81.5 with reasoning-enhanced embeddings, outperforming the strongest baseline by 3.1 points. Its discriminative embeddings also provide a competitive alternative when low-latency retrieval is preferred. Further evaluation on MMEB-v3 suggests that AntEmbed generalizes to a broader range of retrieval settings, including tasks related to tool retrieval, GUI control, and agent memory. As future work, we plan to integrate AntEmbed into agent systems and explore how agent-generated reasoning traces and task context can be incorporated into its reasoning-enhanced retrieval process to better align retrieval with the agent's evolving task needs.

\section*{Acknowledgments}
This work is supported by Ant Group through CCF-Ant Research Fund.

\appendix
\section{Prompt Templates for CoT Generation}
\label{app:CoT_prompts}

For each example, we use a multimodal reasoner to construct a query-side or candidate-side CoT augmentation. Each augmentation contains an evidence-grounded rationale (\texttt{think}) and a concise semantic summary (\texttt{answer}); both are appended to the corresponding original input before encoding. Query prompts express retrieval intent, whereas candidate prompts use a top-$K$ confusion set ($K=5$) to select target-grounded, fine-grained evidence, as described in Sec.~\ref{sec:CoT_Construction}. The output does not mention the reference neighbors or contain facts unsupported by the target.

\subsubsection{Visual-Document Query and Candidate CoT}
\label{app:visdoc_CoT_prompts}

The visual-document subset includes Document-QA, in which a query may
include a document screenshot, and Document-Retrieval, in which a text
query seeks a relevant document page. Query augmentations articulate
the information need. Candidate augmentations use the target document
page together with five nearest-neighbor pages from the same candidate
corpus.

\begin{CoTpromptbox}{Prompt for Visual-Document Query CoT}
\promptheading{System}
You write evidence-grounded query rationales for
visual-document retrieval. Express the information need and the
document evidence needed to satisfy it. Do not use candidate labels or
unseen document content.

\promptheading{Output format}
Return exactly one JSON object:
\begin{lstlisting}[style=CoTpromptstyle]
{"think": "...", "answer": "..."}
\end{lstlisting}

\promptheading{Rules}
\begin{itemize}\setlength{\itemsep}{1pt}
\item For Document-QA, use the supplied screenshot and question to
identify the relevant region, field, table cell, chart element, or
passage. State what evidence is needed; do not introduce an answer
that is absent from the visible screenshot and query.
\item For Document-Retrieval, identify the textual anchors and the
page content or layout a relevant document should contain. Describe
required evidence conditionally; do not claim to have inspected a
page that was not supplied.
\item Keep both fields grounded in the supplied query. Preserve
explicit names, dates, quantities, and quotations without inventing
OCR text or numeric values.
\end{itemize}
\end{CoTpromptbox}

\noindent\emph{The system prompt is paired with one of the following
user messages according to the query modality.}

\begin{CoTpromptbox}{User Prompt for Document-QA Queries}
\textbf{Mode:} \texttt{Document-QA}

\textbf{Question:} \texttt{<QUERY>}

\textbf{Attached query document screenshot:}
\texttt{<QUERY\_DOCUMENT\_IMAGE>}

Identify the question's intent and the relevant evidence in the
attached screenshot. Return only the specified JSON object.
\end{CoTpromptbox}

\begin{CoTpromptbox}{User Prompt for Document-Retrieval Queries}
\textbf{Mode:} \texttt{Document-Retrieval}

\textbf{Text query:} \texttt{<QUERY>}

No query document image is supplied. Describe the evidence a relevant
candidate page should contain. Return only the specified JSON object.
\end{CoTpromptbox}

\begin{CoTpromptbox}{Prompt for Visual-Document Candidate CoT}
\promptheading{System}
You generate confusion-aware evidence for one target
document candidate. The output will be appended to that candidate
before embedding. Five similar document candidates are supplied only
to help identify which target-page details are most discriminative.

\promptheading{User input}
\textbf{Target document text, if available:}
\texttt{<TARGET\_DOCUMENT\_TEXT>}

\textbf{Target page image:} \texttt{<TARGET\_PAGE\_IMAGE>}

\textbf{Hard-neighbor pages ($j=1,\ldots,5$):}
\texttt{<NEIGHBOR\_j\_PAGE\_IMAGE>}; optional text
\texttt{<NEIGHBOR\_j\_DOCUMENT\_TEXT>}.

\promptheading{Instructions}
Compare document type, topic, section structure, visible headings,
tables, charts, fields, labels, dates, and numbers. Select the evidence
that best identifies the target among these similar pages. Report only
text legible on, or facts explicitly supplied with, the target page.
Do not transfer a heading, figure, OCR string, or value from a neighbor
to the target. Do not mention the neighbors or the comparison in the
output.

\promptheading{Output format}
Return exactly one JSON object:
\begin{lstlisting}[style=CoTpromptstyle]
{"think": "...", "answer": "..."}
\end{lstlisting}

\promptheading{Field guidance}
Write \texttt{think} as a grounded paragraph about the target's
distinctive document evidence. Write \texttt{answer} as a concise,
standalone description of the target page. Omit uncertain OCR rather
than guessing; avoid query-specific answers and metadata.
\end{CoTpromptbox}

\bibliographystyle{unsrt}  
\bibliography{references}

\end{document}